\documentclass[a4paper,fleqn,usenatbib]{mnras}

\usepackage[T1]{fontenc}
\usepackage{ae,aecompl}

\usepackage{graphicx}	% Including figure files
\usepackage{amsmath}	% Advanced maths commands
\usepackage{amssymb}	% Extra maths symbols
\usepackage{tikz}
\usetikzlibrary{calc,intersections,through}

\title[HII region model of GC scattering]{A single HII region model of
  the strong interstellar scattering towards Sgr A*}

\author[Sicheneder \& Dexter]{
Egid Sicheneder\thanks{E-mail: egid@mpe.mpg.de}
and Jason Dexter\thanks{jdexter@mpe.mpg.de}
\\
Max Planck Institute for Extraterrestrial Physics,
Giessenbachstr. 1, 85748 Garching, Germany\\
}

\date{Accepted XXX. Received YYY; in original form ZZZ}

\pubyear{2015}

\begin{document}
\label{firstpage}
\pagerange{\pageref{firstpage}--\pageref{lastpage}}
\maketitle

% Abstract of the paper
\begin{abstract}
Until recently, the strong interstellar scattering observed towards
the Galactic center (GC) black hole, Sgr A*, was thought to come from dense
gas within the GC region. The pulse broadening towards
the transient magnetar SGR J1745-2900 near Sgr A* has shown that the
source of the scattering is instead located much closer to Earth,
possibly in a nearby spiral arm. We show that a single
HII region along the line of sight, $1.5-4.8$ kpc away from Earth with
density $n_e$ of a few $\simeq 100 \hspace{2pt} \rm cm^{-3}$ and radius $R
\simeq 1.8 - 3.2$ pc can explain the observed angular broadening of
Sgr A*. Clouds closer to the GC overproduce the observed DM, providing
an independent location constraint that agrees with that from the
magnetar pulse broadening. Our model predicts that sources within
$\lesssim 10$ pc should show the same scattering
origin as the magnetar and Sgr A*, while the nearest known pulsars with
separations $> 20$ pc should not. The radio spectrum of Sgr A* should
show a cutoff from free-free absorption at $0.2 \lesssim \nu \lesssim 1$
GHz. For a magnetic field strength $B \simeq  15 - 70 \hspace{2pt}
\rm \mu$G, the HII region could produce the rotation measure of the magnetar, the largest of any
known pulsar, without requiring the gas near Sgr A* to be strongly magnetised. 
\end{abstract}

% Select between one and six entries from the list of approved keywords.
% Don't make up new ones.

%%%%%%%%%%%%%%%%%%%%%%%%%%%%%%%%%%%%%%%%%%%%%%%%%%

%%%%%%%%%%%%%%%%% BODY OF PAPER %%%%%%%%%%%%%%%%%%

\section{Introduction}
Interstellar scattering by electron density fluctuations along the
line of sight blurs radio images and pulsar emission profiles (angular
and temporal broadening). Certain
lines of sight through the Galaxy show anomalously strong
scattering, notably towards the Galactic center (GC) black hole, Sgr
A*, whose image has been broadened to a constant 1 GHz size of $\simeq 1$
arcsec over $\simeq$ 40 years of observations
\citep[e.g.,][]{backer1978,krichbaumetal1993,loetal1993,loetal1998,shenetal2005,boweretal2006,boweretal2014scat}. 

The origin of the strong scattering towards the GC remains uncertain. From
the lack of free-free absorption of the Sgr A* spectrum,
\citet{vanlangeveldeetal1992} argued that the scattering source should
be located $\Delta \gtrsim 200$ pc from Sgr A*. From a decrease in the number
density of extragalactic background sources near Sgr A*,
\citet{laziocordes1998} found a best fit location of $\Delta \simeq
50-300$ pc. Producing the large observed image of Sgr A* from turbulent gas
so close to the GC would require either extreme turbulent energy
densities or a special scattering geometry \citep{lithwickphd,goldreichsridhar2006}.

A prediction of this scattering model was that radio pulsars in the GC
should be rendered undetectable due to the spread of arrival times of their
pulses becoming longer than their pulse periods. Recently, a rare
transient magnetar SGR J1745-2900 was discovered $\simeq 0.1$ pc
($\simeq 2.5$ arcsec) in projection from Sgr A* \citep{kenneaetal2013,morietal2013}. Radio pulsations were
detected from this source down to $1$ GHz \citep{eatoughetal2013}, and
the pulse broadening from scattering was measured to be $2-3$ orders
of magnitude smaller than predicted \citep{spitleretal2014}.  In
addition, the elliptical scatter-broadened image of J1745-2900 was found to be identical to that of Sgr A* in
both size and orientation \citep{boweretal2014}. The combination of
angular and temporal pulse broadening from the same object allowed an
estimate of the location of the scattering along the line of sight,
$\Delta = 5.9 \pm 0.3$ kpc, far from the GC. The chance alignment
of such a scattering region, of size $\lesssim 0.5$ deg \citep{vanlangeveldeetal1992}, with Sgr A* and the GC is strange unless
strong scattering has a common astrophysical origin.

HII regions have long been thought to produce strong interstellar
scattering towards the Galactic plane
\citep[e.g.,][]{litvak1971,little1973,dennisonetal1984}. Here we assess the
physical conditions required to produce the 
observed scattering towards Sgr A* in terms of a simple model of turbulent,
ionized gas in such an HII region (\S \ref{sec:Methods}). We show that
the model can explain the
observed scattering properties for a typical size and electron density
(\S \ref{sec:Results}). From the range
of allowed cloud properties (\S \ref{sec:Discussion}), we calculate
its contribution to the RM of the magnetar and make
predictions for the scattering locations of other GC pulsars and the
low-frequency cutoff to the radio spectrum of 
Sgr A*. 

\section{Methods}\label{sec:Methods}

We use the thin screen approximation
\citep[e.g.,][]{ishimaru1977,blandfordnarayan1985,vanlangeveldeetal1992} to calculate the angular
broadening associated with a given spectrum of density fluctuations
arising from an HII region.

\subsection{Thin screen scattering model}\label{sec:Basics}

Turbulent, free electrons scatter electromagnetic radiation strongly, since the electron density fluctuates. The electron density fluctuations are described by \citep{1985Boriakov},
\begin{align}\label{FlucDef}
\langle \delta n_e^2 \rangle = \int {P_{\delta n_e}}(\mathbf{q}) \,\mathrm{d}\mathbf{q},
\end{align} with the Kolmogorov spectrum,
\begin{align}\label{Kol}
{P_{\delta n_e}}(\mathbf{q})={C_n (x)}^2\cdot \mathbf{q}^{-\frac{11}{3}} ; \qquad \text{for } \mathbf{q}_0 \ll \mathbf{q} \ll \mathbf{q}_1 ,
\end{align}

\noindent where $\mathbf{q}$ denotes the wavenumber, and $q_1 = 2\pi /l_1$ and
$q_0 = 2\pi / l_0$ correspond to the inner and outer scales of the
turbulenct spectrum, $l_1$ and $l_0$. Assuming isotropic turbulence, the visibility $V(\rho;L)$ is given by
\citep{vanlangeveldeetal1992}, 
\begin{align}\label{theory}
V(\mathbf{\rho}; D)&=\exp{\left(-D_{\mathbf{\Phi}}/2\right)}, \\
D_{\mathbf{\Phi}} (\rho)&=8 \pi r_e^2 \lambda^2 \int_0^D \,\mathrm{d}x \int \,\mathrm{d}q \, q \left[1-J_0 (q \rho)\right] {P_{\delta n_e}}(q),
\end{align}
\noindent where $\mathbf{\rho}$ is the baseline length, $D$ is the
distance from the observer to the source and $J_0 (q \rho)$ a Bessel function. The phase structure function
$D_{\mathbf{\Phi}}$ specifies the statistical properties of the
turbulent medium. 

For $\rho < 2\pi/l_1=q_1$, the visibility has a Gaussian profile:

\begin{align}
V(\mathbf\rho;D)&=\exp{\left(-\rho^2/\rho^2_{C}\right)}, \\
\rho_{C}&=\left[6\pi^2 \lambda^2 r_e^2 \mathcal{L}(D) q_1^{\frac{1}{3}}\right]^{-\frac{1}{2}}\label{eq:1},
\end{align}

\noindent as is observed for Sgr A*. Here $r_e$ is the classical
electron radius. The function $\mathcal{L}(D)$ takes into account
the strength and position of the turbulent medium
along the line of sight, 

\begin{align}\label{eq:RHO}
\mathcal{L}(D)&=\int_0^D C_n^2 (x)  \left(\frac{x}{D}\right)^{2}
                \,\mathrm{d}x \simeq C_n^2 \gamma^2 H,
\end{align}

\noindent where $x$ is the distance of the screen from the
  source and $\gamma \equiv \Delta / D$ the relative screen
  location. The second step comes from taking $C_n^2 (x)$ to be
  constant across the HII region and assuming the cloud thickness $H$ is small 
($H \ll D$).

The apparent image size
$\theta$ and $\rho_C$ are related by, 

\begin{align}\label{EssencialEq}
\rho_C=\frac{\lambda \cdot \sqrt{2 \ln (2)}}{\pi \cdot \theta}.
\end{align}

\noindent We express the structure constant in
terms of the outer scale $l_0$ \citep{1985Boriakov},

\begin{align}\label{eq:2}
C_n^2=\frac{\delta n_e^2}{6\pi}\left(\frac{2\pi}{l_0}\right)^{2/3}.
\end{align}

\noindent Using the thin screen approximation and equations
\ref{eq:1}-\ref{eq:2}, we can write the image size as,

\begin{align}\label{eq:3}
\theta = \frac{2 \sqrt{\ln 2} \hspace{2pt} r_e \lambda^2 
  \hspace{2pt} \delta n_e \gamma H^{1/2}}{l_0^{1/3} l_1^{1/6}}.
\end{align}

\subsection{An HII region as a thin scattering screen }\label{sec:HII}
We consider a uniform distribution of free electrons contained in a
single HII region, with a size given by its Str\"{o}mgren radius
$R_S$,

\begin{align}\label{Stroe}
R_S=\left(\frac{3N_{Ly}}{4 \pi \alpha_H n_{e}^{2}}\right)^{\frac{1}{3}}\simeq 4.7 \left(\frac{n_e}{100\, \rm cm^{-3}}\right)^{-2/3}\cdot N_{Ly,f}^{1/3}\hspace{2pt} \rm pc,
\end{align}

\noindent for an ionizing photon rate $N_{Ly} = 5 
\times 10^{49} N_{Ly,f} \hspace{2pt} \rm s^{-1}$, scaled to a value appropriate
for a bright O star. The scaling factor $N_{Ly,f}$ is a model
parameter, with $N_{Ly,f} = 1$ in the fiducial case. At the radius $R_S$, the photoionization is in
equilibrium with the recombination, characterized by the volumetric
recombination rate $\alpha_H \simeq 4\times10^{-13} \hspace{2pt} \rm
cm^{-3} \, s^{-1}$.

We connect the properties of the turbulent fluctuations with the
density and size of the HII radius by assuming the line of sight
passes through a cloud thickness $H = 4/3 R_S$, the average value for a
line of sight through a sphere, and parameterize the fluctuations as
$\delta n_e = f n_e$, with $f \le 1$. We
further assume an inner scale $l_1 = 10^4$ km
\citep[e.g.,][]{wilkinsonetal1994} and an outer scale 
comparable to the cloud radius: $l_0 = f_2 R_S$ with $f_2 \le 1$. With
these scalings, we re-write equation \ref{eq:3} as, 

\begin{equation}\label{eq:theta}
\theta \simeq 0.4 \,
  \gamma \, f_2^{-1/3} \, f \left(\frac{n_e}{100 \rm \, cm^{-3}}\right) \left(\frac{R_S}{1\, 
        \rm pc}\right)^{1/6} \left(\frac{\lambda}{30 \, \rm cm}\right)^2
    \rm arcsec.
\end{equation}

\noindent where $\gamma = \Delta/D$ is the relative screen location
along the line of sight defined in terms of the screen-source
($\Delta$) and total ($D$) distance. The same cloud properties produce a larger image when
located closer to the observer.

\subsection{Geometrical implications}

Due to scattering, different light rays reach the observer at different
times. The width of the resulting broadened pulse depends on the image size and the thin
screen location as\citep[e.g.,][]{cordeslazio1997},

%\textcolor{red}{Taylor!}
\begin{align}
t_{\rm PB}\simeq \frac{D}{8\cdot ln(2)\cdot c}\frac{1-\gamma}{\gamma}\cdot \theta^2 \hspace{2pt} \rm sec,
\label{eq:PulseTimeDelay}
\end{align} called pulse broadening.

With a cloud size $R_s$ and location $\gamma$, our model gives the
mean angular separation $\theta_{\rm sep}$ of two sources, such that they
can not be scattered by the same cloud:

\begin{align}\label{eq:angular_separation}
\theta_{\rm sep} &\simeq
               \frac{R_S}{D-\Delta}=\frac{R_S}{D}\frac{1}{1-\gamma},\\
\theta_{\rm sep} &\simeq 0.033\cdot
               \frac{N_{Ly,f}^{\frac{1}{3}}}{1-\gamma}
               \left(\frac{n_e}{100 \,
               \rm cm^{-3}}\right)^{-\frac{2}{3}}\hspace{2pt} \rm deg.
\end{align}

Second, the Earth rotates around the Galactic center at a speed
$v_{Orb} \simeq 220$ km/s. We can
calculate in our framework how long it takes for the earth to pass the
cloud, changing the observed scattering properties:

\begin{align}
t_{\rm pass}&=\frac{R_{S}}{v_{Orb}} \frac{D}{D-\Delta} = 
\frac{R_{S}}{v_{\rm orb}}  \frac{1}{\gamma},\\
t_{\rm pass}&=21000 \cdot \frac{N_{Ly,f}^{\frac{1}{3}}}{\gamma} \left(\frac{n_e}{100 \, \rm cm^{-3}}\right)^{-\frac{2}{3}} 
\hspace{2pt}
\rm yr.
\end{align}

\begin{figure}
\centering
    \includegraphics[width=0.5\textwidth]{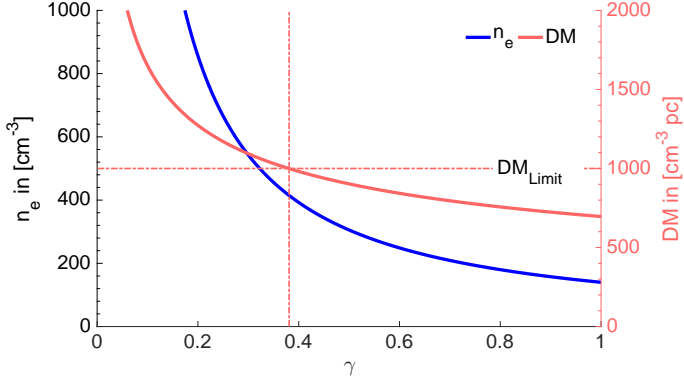} 
  \caption{\label{fig:density}To account for the observed angular broadening of Sgr A*, the electron density (blue line and left axis) and
    dispersion measure (red line and right axis) must increase with
    decreasing distance between the model HII region and Galactic
    center. The horizontal line marks our assumed upper limit on the
    contribution to the dispersion
    measure of J1745-2900 from the HII region, $\rm DM_{\rm Limit}=1000 \rm pc \, \,
    cm^{-3}$, in turn constraining the screen location $\gamma \gtrsim
    0.4$.}
\end{figure}

\begin{figure}
  \centering
    \includegraphics[width=0.5\textwidth]{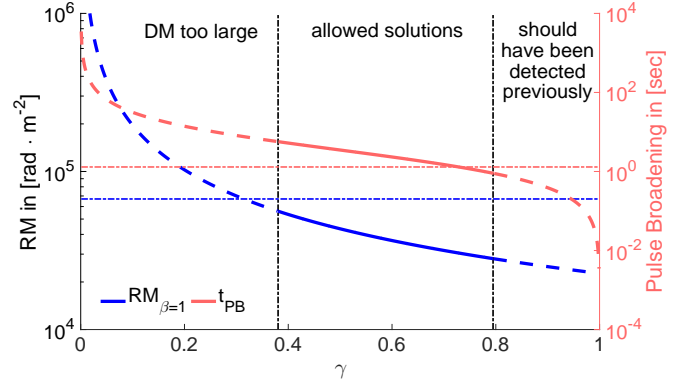} 
  \caption{ \label{fig:RMTAU}Rotation measure (RM, blue line and left
    axis) and pulse broadening (red line and right axis) versus
      the relative position of the HII region. Solid lines show the
      allowed solutions and the horizontal lines show the limits for DM and the pulse broadening. For
      clouds close to the GC ($\gamma \lesssim 0.4$), the required densities become large,
      overproducing the observed DM and $\tau_{\rm ff}$. A cloud 
      close to Earth ($\gamma \gtrsim 0.8$) likely would have 
      been detected already and so is not considered viable.}
 \end{figure}

\section{Results}\label{sec:Results}

We calculate the electron density $n_e$ and corresponding radius $R_S$
required to produce the observed angular broadening of Sgr A* and
J1745-2900 \citep[$\theta = 945$ mas at 1 GHz,][]{boweretal2014} from
equation \eqref{eq:theta} as a function of the screen
location $\gamma$ along the line of sight:

\begin{align}\label{eq:Case2}
n_e &\simeq 140 \cdot f_2^{\frac{3}{8}}\cdot f^{-\frac{9}{8}}\cdot
\gamma^{-\frac{9}{8}}\cdot N_{Ly,f}^{-1/16}
\hspace{2pt} \rm cm^{-3},\\
R_S &\simeq 3.7 \cdot f^{\frac{3}{4}}\cdot f_2^{-\frac{1}{4}}\cdot
      N_{Ly,f}^{3/8}\cdot \gamma^{\frac{3}{4}}
      \hspace{2pt} \rm pc,
\end{align}

From the cloud density and radius, we calculate its
contribution to the magnetar dispersion and rotation measures:

\begin{align}
\rm DM &= \int_0^L n_e (x) \,\mathrm{d}x \nonumber\\
&\simeq 620 \left(\frac{n_e}{100\, \rm cm^{-3}}\right)^{1/3}
\cdot N_{Ly,f}^{1/3} \hspace{2pt} \rm pc \, \, cm^{-3},\label{eq:DM}\\
\rm RM &= 0.81 \int_0^L n_e (x) \cdot \mathbf{B} \,\mathrm{d}\mathbf{x}\nonumber\\
&\simeq 5000
\cdot \left(\frac{n_e}{100\, \rm cm^{-3}}\right)^{1/3}\cdot
\left(\frac{B_{||}}{10\, \rm \mu \rm G}\right)\cdot N_{Ly,f}^{1/3} \hspace{2pt} \rm rad\, \, m^{-2}\label{eq:RM}\\
B_{||} &\simeq 34\cdot \beta^{-\frac{1}{2}}\cdot T_{e,4}^{\frac{1}{2}}\cdot\left(\frac{n_e}{100\, \rm cm^{-3}} \right)^{1/2}\,  \rm \mu \rm G \label{eq:BField}
\end{align}

\noindent Using equation \eqref{eq:Case2}, we can re-write these in
terms of the location $\gamma$ of the cloud:

\begin{align}
\rm DM &\simeq 700\cdot f_2^{\frac{1}{8}}\cdot f^{-\frac{3}{8}}\cdot
\gamma^{-\frac{3}{8}}
\cdot N_{Ly,f}^{5/16}
 \hspace{2pt} \rm pc \, \, cm^{-3},\label{eq:DMpara}\\
\rm RM &\simeq 22,700 \cdot \beta^{-1/2}\cdot T_{e,4}^{1/2 }\cdot f_2^{\frac{5}{16}}\cdot f^{\frac{15}{16}}
\cdot \gamma^{-\frac{15}{16}} \cdot N_{Ly,f}^{9/32}\cdot 
%q_{1,f}^{-5/32} 
\hspace{2pt}
\rm rad\, \, m^{-2},\label{eq:RMpara}
\end{align}

\noindent where $\beta$ is the ratio of the thermal to magnetic 
pressure and $T_{e,4} = T_e / 10^4 \, K$ is the electron
temperature. We set the parameters $f$, $f_2$ and $N_{Ly,f}$
  (introduced in section \S \ref{sec:HII}), equal to 1, unless
  otherwise mentioned.  Figure \ref{fig:density} shows the particle
density $n_e$ and $\rm DM$ as a function of the screen location. Both quantities increase for screens
closer to the GC, where producing the observed angular size requires
large values of $C_n^2$.

\noindent We also calculate the free-free optical depth $\tau_{\rm ff}$ ~\citep{rybicki1979radiative} through
the cloud:

\begin{align}
\tau_{\rm ff} &\simeq 0.024 \cdot T_{e,4}^{-3/2} \cdot
            N_{Ly,f}^{\frac{1}{3}}\cdot \left(\frac{n_e}{100
            \, \rm cm^{-3}}\right)^{4/3} \left(\frac{\nu}{1
            \, \rm GHz}\right)^{-2}\nonumber\\
&\simeq 0.038 \cdot f_2^{\frac{1}{2}}\cdot f^{-\frac{3}{2}} \cdot
  T_{e,4}^{-\frac{3}{2}} \gamma^{-\frac{3}{2}} N_{Ly,f}^{1/4},
 \label{eq:tffpara}
\end{align}

\noindent where we use $\nu=1$ GHz and set the Gaunt factor $g_{\rm ff} \simeq 7$
for this frequency \citep{karzaslatter1961}. Finally, the 
geometric quantities are given as,

\begin{align}
\theta_{\rm sep}\simeq 0.027 \cdot f_2^{-\frac{1}{4}}\cdot f^{\frac{3}{4}}\cdot
\frac{\gamma^{\frac{3}{4}}}{1-\gamma}\cdot
  N_{Ly,f}^{\frac{3}{8}}
\hspace{2pt} \rm deg,\\
t_{\rm pass}=16,800 \cdot f^{\frac{3}{4}}\cdot f_2^{-\frac{1}{4}}\cdot
  N_{Ly,f}^{3/8}\cdot \frac{\gamma^{\frac{3}{4}}}{1-\gamma} \hspace{2pt} \rm yr. 
\end{align}

We compare these model values to the measured quantities from the line of sight
towards Sgr A* / J1745-2900. We use upper limits of $ \rm DM < 1000
\rm \hspace{2pt} pc \rm \hspace{2pt} cm^{-3}$,
$\tau_{\rm ff} < 1$ at 1 GHz \citep{meliafalcke2001,royetal2004}, and
lower limits of $t_{\rm pass} > 40$ yr, $\theta_{\rm sep} > 2.5$ arcsec to constrain the allowed range
of screen locations. This DM limit is smaller than the total value
towards J1745-2900 \citep[$ \rm DM = 1778 \rm \hspace{2pt} pc \rm
\hspace{2pt} cm^{-3}$][]{eatoughetal2013}. We use it as a limit
because it is comparable to both the observed DM values toward the nearest pulsars to the GC
\citep{johnstonetal2006,denevaetal2009detect}, and to the Galactic
disc component of the DM along this line of sight in the
NE2001 model \citep[e.g., the $\Delta \rm DM$ between lines of sight with
 $l = 0^\circ$ and $l = 2^\circ$ with $b = 0^\circ$,][]{ne2001}.

Screens with $\gamma \lesssim 0.4$ exceed this DM limit, and so are
excluded. A weaker limit $\gamma \gtrsim 0.1$ comes from the free-free
optical depth towards Sgr A*. For all models, the time
for the line of sight to move across the cloud is much longer than the
$40$ years over which the size of Sgr A* has been measured, and so we
do not consider this constraint further. These constraints do not rule
out large HII regions very close to Earth. Nearby HII regions should already
have been detected, and so we restrict the allowed models to those
with $D-\Delta > 1.5$ kpc.

Figure \ref{fig:RMTAU} shows the temporal broadening (red curve, right
axis scale) associated with
our thin screen model as a function of
its location (equation \ref{eq:PulseTimeDelay}). \citet{boweretal2014}
combined angular and temporal broadening measurements of J1745-2900
\citep{spitleretal2014} to locate the screen location as $\Delta = 5.9
\pm 0.3$ kpc. Our constraints are independent of and consistent with
this measurement of the scattering location.

\begin{figure}
  \centering
    \includegraphics[width=0.5\textwidth]{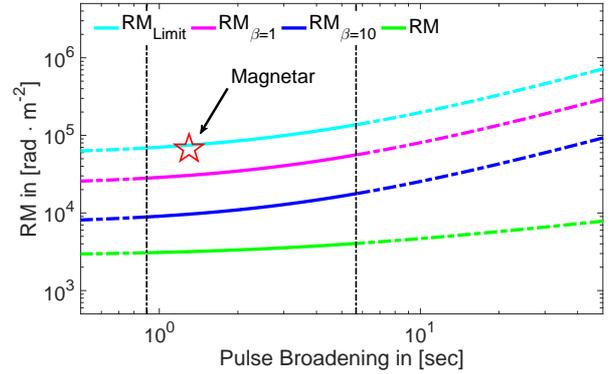} 
  \caption{\label{fig:RMvsTau}Rotation measure (RM) versus the pulse broadening. A magnetised cloud (cyan and purple lines) can
    produce the observed $ \rm RM \simeq 6.7 \cdot 10^4  \hspace{2pt} \rm rad\, \, m^{-2}$ (red star, see \citep{eatoughetal2013}) which was previously thought to
    require a dynamically important magnetic field in gas falling onto
    Sgr A* \citep{eatoughetal2013}.}
\end{figure}

The left hand axis and blue curve in figure \ref{fig:RMTAU}
show the model RM for an equipartition strength magnetic field
($\beta=1$). At $\gamma \simeq 0.4$ the
model (see equation \eqref{eq:RM}) can explain even all of the observed $ \rm RM \simeq 6.7 \cdot 10^4  \hspace{2pt} \rm rad\, \, m^{-2}$ \citep{eatoughetal2013} towards
J1745-2900, the largest of any pulsar. The magnetic field strength at $\gamma \simeq 0.4$ is $B_{||} \simeq  70 \rm \hspace{2pt} \mu G$ (see equation \eqref{eq:BField} with $\beta=1$). Towards $\gamma \simeq 0.8$ the model RM and magnetic field strength drop, at least able to explain the bulk of the observed RM towards J1745-2900. Where the magnetic field strength is $B_{||} \simeq  45 \rm \hspace{2pt} \mu G$. This is further shown in figure \ref{fig:RMvsTau}, which shows $\rm RM$ vs. $t_{\rm PTD}$ for
our model with different assumed field strengths compared to the
observed values from J1745-2900. Except for very low field strengths \citep[e.g.,][]{harveysmithetal2011},
the allowed models contribute significantly to the observed RM.

\begin{figure}
  \centering
    \includegraphics[width=0.5\textwidth]{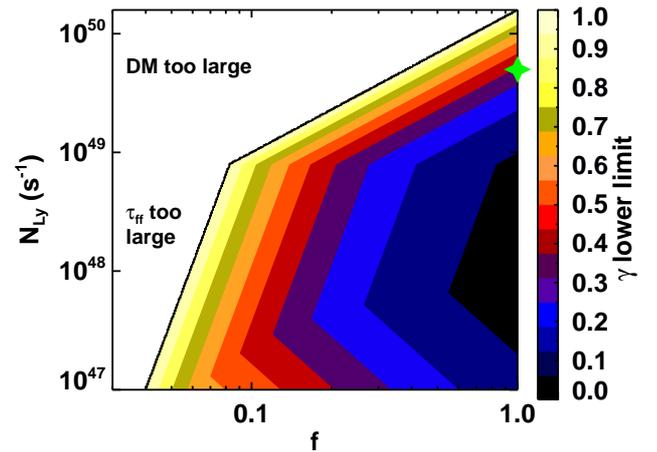} 
  \caption{\label{fig:gamma}Lower limit on screen distance from the GC
    ($\gamma$) as a
      function of the ionizing photon rate $N_{Ly}$ and fluctuation
      strength $f$. The relevant constraints are $\rm DM < 1000 \rm \hspace{2pt} pc \rm \hspace{2pt}
cm^{-3}$, $\tau_{\rm ff} < 1$ at 1 GHz, and $\theta_{\rm sep} > 2.5$
      arcsec. The green star shows our fiducial parameter choices
      $f=1$, $N_{Ly} = 5\times10^{49} \hspace{2pt} \rm s^{-1}$. At large $N_{\rm Ly}$, low
    $f$ models overproduce the observed DM, while at low $N_{\rm Ly}$
    the optical depth is too large. In the lower right region, the
    limit on $\gamma$ comes from forcing the cloud to be large enough
    to produce both the magnetar and Sgr A* images. The
      white region shows models where all $\gamma$ are excluded,
      placing a constraint $f \gtrsim 0.1$. The range of allowed
      screen locations we find is generic to a large part of the
      parameter space. Screens close to the GC ($\gamma \gtrsim 0.05$)
      would require a weak ionizing source driving strong
      turbulence, and are only possible in a narrow range of
      parameter space.}
\end{figure}

\begin{figure}
  \centering
    \includegraphics[width=0.5\textwidth]{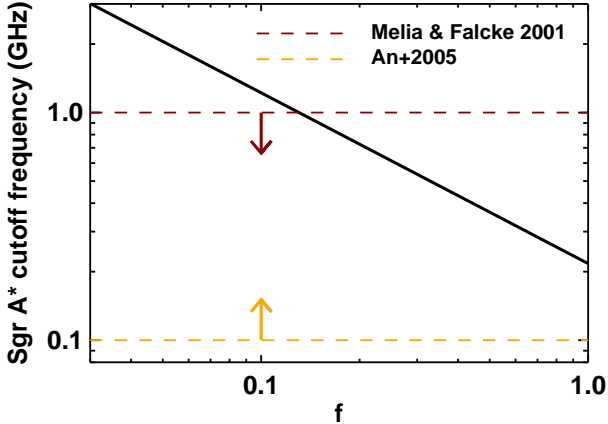} 
  \caption{\label{fig:taucutoff}Predicted cutoff frequency (where $\tau_{\rm ff} = 1$) for the radio spectrum of Sgr A*
  as a function of the fluctuation strength $f$, assuming
  a screen location $\gamma=0.75$ found by
  \citet{boweretal2014}. Detections of Sgr A* below 1 GHz constrain $f
\gtrsim 0.1$, while our model predicts a cutoff should be seen at a
frequency above $\simeq 200$ MHz, in agreement with low frequency
spectral measurements \citep{nordetal2004,anetal2005}.}
\end{figure}

\section{Discussion}\label{sec:Discussion}

The recently discovered GC magnetar SGR J1745-2900, $\simeq 0.1$ pc
from Sgr A* in projection, shows the same scatter-broadened radio
image as Sgr A*, but pulse broadening $2-3$ orders of magnitude
smaller than predicted \citep{boweretal2014,spitleretal2014}. The
combination of these measurements implies that the turbulent gas
producing the observed image is located far from the GC.

HII regions have long been candidates for the observed strong
interstellar scattering towards the Galactic plane. We have shown that
for typical properties, $n_e \simeq 100\,  \rm cm^{-3}$, $R_S \simeq
1.8 - 3.2$ pc, such an HII
region $1.5-4.8$ kpc from Earth can explain the angular broadening of Sgr A*. Placing the
screen closer to the GC ($\gamma \lesssim 0.4$) overproduces the
observed DM of the magnetar, and for screens close to the GC ($\gamma
\lesssim 0.1$) also the free-free optical depth towards Sgr A*. This
constraint on the location of the scattering medium is independent of,
and consistent with, the geometric result $\gamma \simeq 0.75$ found
by combining the angular and temporal broadening of the magnetar
\citep[equation \ref{eq:PulseTimeDelay},][]{boweretal2014}.

We have assumed a uniform HII region
with a size and particle density related by the Str\"{o}mgren radius
for an assumed rate of ionizing photons, $N_{Ly} = 5\times10^{49}
\hspace{2pt} \rm s^{-1}$, and fluctuation strength $\delta n_e = f
n_e$ with $f=1$. Figure \ref{fig:gamma} shows how our lower limit
on the screen location, $\gamma$, depends on these
parameters. Decreasing the turbulent scaling, $f$, increases the
required particle density to produce the angular broadening of Sgr
A*. This in turn increases the DM and $\tau_{\rm ff}$ at each location, and so requires
the screen to be located farther from the GC. A similar effect results
from increasing $N_{Ly}$. This causes the HII region to be
larger for fixed $\delta n_e$, and increases the DM since its
weighting with $R$ is stronger than that of $C_n^2$. The
combination of these effects means that the model is only compatible
with the observed DM and $\tau_{\rm ff}$ ($\gamma < 1$) for strong turbulence, $f
\gtrsim 0.1$, and $N_{Ly} 
\lesssim 10^{50} \hspace{2pt} \rm s^{-1}$. In models of MHD
turbulence, it is sometimes assumed that $f \simeq \beta^{-1}$
\citep[e.g.,][]{goldreichsridhar2006}. Our limit on $f$ could then
favor a relatively large value of $B$, which in turn leads to a larger
contribution of the HII region to the observed RM.

A successful model of scattering towards the magnetar also needs to
account for its temporal broadening, $1.3 \pm 0.2$s at 1 GHz \citep{spitleretal2014}, which fixes $\gamma \simeq 0.75$
\citep{boweretal2014} in a thin screen model. For this location, we find the following HII
region properties: $n_e \simeq 200 \hspace{2pt} \rm cm^{-3}$, $R_S \simeq 3$ pc, $\rm DM
\simeq 800 \hspace{2pt} \rm pc \, \,
    cm^{-3}$, $\tau_{\rm ff} \simeq 0.05 \hspace{2pt} \nu_{\rm GHz}^{-2}$,
$\theta_{\rm sep} \simeq 5$ arcmin. The nearest pulsars in
angular separation are $10-15$ arcminutes from Sgr A*
\citep[$\simeq 25$ pc at the
distance of the GC,][]{johnstonetal2006,denevaetal2009detect}. All of these GC pulsars
show large pulse broadening, and so measurements of
their radio images can constrain the screen location $\gamma$ in the
same fashion as done by \citet{boweretal2014} for SGR J1745-2900. A
prediction of our model is that the single HII region is unlikely to
cover all of these pulsars, so that their values of $\gamma$ should be
different than that of the magnetar. This prediction is consistent
with recent measurements of angular broadening for the other GC
pulsars (Dexter et al., in prep.). Significant angular broadening of
OH/IR stars is seen on larger scales $\lesssim 0.5$ deg
\citep[][]{vanlangeveldeetal1992,frailetal1994}. In
our model there would need to be multiple clouds covering the
region. This is consistent with the variation of the maser angular sizes and the
free-free optical depth to extragalactic background sources on these
scales \citep{roy2013}.

Our model also predicts a low-frequency cutoff to the
Sgr A* spectrum ($\tau_{\rm ff} = 1$) at $\gtrsim 0.2$ GHz (Figure
\ref{fig:taucutoff}. The smallest values of the cutoff frequency occur
for $f = 1$, while values $f \lesssim 0.1$ are ruled out by the
unbroken power law Sgr A* spectrum down to $1$ GHz
\citep{meliafalcke2001}, showing again that strong turbulence is
needed to explain the scattering towards Sgr A*. \citet{anetal2005}
and \citet{nordetal2004} find evidence for a break in the spectrum at
wavelengths between $47-100$ cm ($100-866$ MHz), consistent with our
prediction. Measuring the low-frequency cut off shape of the Sgr A*
spectrum would help to directly measure the fluctuation strength and
further constrain the model.

The RM of the magnetar is an order of magnitude larger than that of
any other pulsar, and for this reason was previously thought to come from gas local
to the GC. To produce the observed RM, this gas would need to be
threaded by very strong, uniform magnetic fields
\citep{eatoughetal2013}. As an alternative, we show that for magnetic
field strengths $\simeq 15-70\, \mu$G a single HII region can produce
much or all of the observed RM (figure \ref{fig:RMvsTau}). This field
strength is large, but within the range of observed HII region values
of both B \citep[e.g.,][]{heilesetal1981,rodriguezetal2012} and $\beta$
\citep{harveysmithetal2011}. Therefore we caution that the RM of the magnetar
does not necessarily require that the gas near Sgr A* be highly
magnetised. The HII region cannot however produce the order of magnitude larger RM
seen towards Sgr A* itself, which is thought to arise within the
surrounding accretion flow \citep{boweretal2003,marroneetal2007}.

\citet{schnitzeleretal2016} measured the RM towards the other GC pulsars and found two others with
$\rm RM \simeq 10^4 \hspace{2pt} \textrm rad \hspace{2pt} \textrm m^{-2}$. If the very large RM for SGR
J1745-2900 is produced from extremely strong, ordered fields within
the central parsec, it seems strange that smaller but comparable RMs would be
found for these other objects much further away. The
mean field strength estimated from $\rm RM/ \rm DM \simeq 30\, \mu$G for these pulsars
is similar to that of SGR J1745-2900, and so distant HII regions with
mean field strengths like we require could be a more natural
explanation. On the other hand, the RM towards the magnetar and other
GC pulsars is an order of magnitude higher than for other known
pulsars, while our model should apply to many heavily scattered lines
of sight in the inner Galaxy. This suggests that other heavily
scattered lines of sight either have weaker field strengths (e.g. the
HII regions towards the GC would have to be uncomfortably ``special''), are
preferentially not detected by pulsar surveys (preventing detections
of large RMs away from the GC), or that the GC environment on scales
of tens of pc does in fact produce the large observed RMs as suggested
by \citet{schnitzeleretal2016}.

In this scenario for the scattering towards Sgr A*, the small HII region is
aligned with Sgr A* by chance and does not cover the entire GC. The
chance probability of this occurrence is small unless lines of sight with such strong
scattering are common. A significant fraction ($\gtrsim 10\%$) of lines of sight
through the inner Galaxy pass through a known HII region \citep{andersonetal2014}, and
gas with typical densities we find of $\sim 100 \hspace{2pt} \rm cm^{-3}$ is seen 
strongly in emission towards the inner Galaxy at the radial velocity corresponding to the Scutum
spiral arm \citep{langeretal2016}. A handful of lines of sight
with very strong scattering are known \citep[e.g.,][]{rodriguezetal1982,wilkinsonetal1994} and many extragalactic
background sources behind the Galactic plane are known to be heavily
scattered
\citep[e.g.,][]{lazioetal1999,claussenetal2002,beasleyetal2002,pushkarevkovalev2015}. If
such lines of sight are common, then HII regions as modeled here 
should contribute significantly to the observed DM and RM towards
heavily scattered objects. Lower limits on the DM, and therefore
revised pulsar distance estimates, can be inferred from 
our model in cases where the properties of the intervening HII region can be measured.

\section*{acknowledgements}
We thank E. Quataert, F. Eisenhauer, S. Gillessen, R. Herrera-Camus,
G. Bower, and R. Wharton for useful discussions. This work was supported by a Sofja
Kovalevskaja Award from the Alexander von Humboldt Foundation of Germany. 

\bibliographystyle{mnras}

% Don't change these lines
\bsp	% typesetting comment
\label{lastpage}

\end{document}